# Phonon anomalies, anharmonicity, and thermal expansion coefficient in few layered $PtX_2$ (X= S, Se): A temperature dependent Raman study

Atul G. Chakkar[1,*], Chaitanya B. Auti[1], Gaurav Bassi[2], Mukesh Kumar[2], and Pradeep Kumar[1,#]

[1]*School of Physical Sciences, Indian Institute of Technology Mandi, Mandi-175005, India*

[2]*Functional and Renewable Energy Materials (FREM) Laboratory, Department of Physics, Indian Institute of Technology Ropar, Rupnagar, Punjab, 140001, India*

**E-mail: atulchakkar16@gmail.com*

*#E-mail: pkumar@iitmandi.ac.in*

**Abstract:**

Two-dimensional group-10 noble transition metal dichalcogenides have garnered growing attention due to their rich physical properties and promising applications across nanoelectronics, optoelectronics, and spintronics. Among them, $PtX_2$ (X = S, Se) exhibits pronounced interlayer coupling driven by hybridization of the out-of-plane $P_z$ orbitals of the chalcogen atoms. In this work, we present a detailed temperature and polarization-resolved Raman spectroscopic study of few-layer $PtS_2$ and $PtSe_2$ over the temperature range of ~ 5 to 300 K. Our study encompasses phonon-phonon interactions, symmetry analysis of phonon modes, low-frequency interlayer vibrations, and extraction of thermal expansion coefficients. Notable phonon anomalies in peak position, linewidth, and intensity emerge around ~ 80 K and 150 K for $PtS_2$, and ~ 70 K and 240 K for $PtSe_2$, indicating intricate coupling between thermal and vibrational dynamics. These results offer valuable insights for the development of devices based on $PtS_2$, $PtSe_2$, and related 2D materials, where interlayer interactions, anharmonic effects, and thermal expansion behaviour play crucial roles.

## 1. Introduction

Over the past decade, two-dimensional transition metal dichalcogenides (TMDCs) have garnered significant attention due to their remarkable optical band gaps, diverse electronic, thermal, and vibrational properties, and broad applicability in optoelectronic, nanoelectronics, and device fabrication technologies. TMDCs typically follow the formula $MX_2$, where M is a transition metal and X is a chalcogen (S, Se, or Te). Among these, group-6 TMDCs such as $MoS_2$, $MoSe_2$, $WS_2$, and $WSe_2$ are extensively investigated for their tunable band gaps, rich excitonic behaviour [1-13], and wide-ranging applications [14-19]. Recent studies on group-10 noble TMDCs (TM = Ni, Pd, Pt; X = S, Se, Te) have revealed phenomena including superconductivity, intriguing optical, magnetic, and thermoelectric properties [20-34], enabling their potential in sensing, optoelectronics, valleytronics, spintronics [25, 26, 39-48], and energy devices [49-51]. Among these compounds, $PtX_2$ (X = S, Se) stands out owing to its strong interlayer coupling, tunable band gap, high carrier mobility, robust air stability, and other exotic features [43, 53-57]. $PtS_2$ is identified as an indirect band gap semiconductor [54], while $PtSe_2$ behaves as a type-II Dirac semimetal [58,59]. Both exhibits pronounced layer-dependent band gaps: $PtS_2$ transitions from ~0.25 eV in bulk to ~1.6 eV in monolayer [53,54]; while $PtSe_2$ is semiconducting (~1.6 eV) in monolayer and semi metallic in bulk [23,43,54], driven by *Pz* orbital hybridization between interlayer S and Se atoms. For $PtSe_2$, the shift from semimetal to semiconductor with reduced thickness down to bi- or monolayers suggests its viability in electronic and optoelectronic applications [60-62]. Given these attributes, understanding interlayer interactions, phonon dynamics, and thermal behaviour becomes essential. Raman spectroscopy has been proved to be an indispensable tool for probing interlayer coupling, thermal expansion, quasi-particle interactions, and layer-dependent characteristics in TMDCs and related 2D materials [63-70].

We conducted a comprehensive temperature-dependent Raman spectroscopic study on few-layer $PtX_2$ (X = S and Se) across the temperature range of 5-300 K using 633 nm laser excitation, aiming to elucidate phonon-phonon interactions, interlayer coupling, thermal expansion effects, and other underlying phenomena. Additionally, polarization-resolved Raman measurements were carried out at room temperature for both compounds to investigate the symmetry of the observed phonon modes.

## 2. Experimental details

Few-layer thin films of 2D $PtS_2$ and $PtSe_2$ were synthesized on $SiO_2$/Si substrates via the thermally assisted conversion (TAC) technique [71,72]. Temperature-dependent Raman scattering measurements were done using a LabRam HR-Evolution Raman spectrometer in the backscattering geometry, with 1800 lines/mm grating. The temperature was precisely regulated between 5 and 300 K using a closed-cycle cryostat (Montana Instruments). A 633 nm laser served as an excitation source, with its power maintained below 0.3 mW to minimize the local sample heating. A 50× (NA = 0.5) long working distance objective lens was employed for both focusing on the sample surface and collecting the scattered signal.

## 3. Results and discussion

### 3.1. Lattice Vibrations

Both bulk and monolayer forms of $PtS_2$ and $PtSe_2$ crystallize in the $CdI_2$-type lattice configuration within the 1T-phase, exhibiting $D_{3d}$ point group symmetry (space group: $P\bar{3}m1$ (#164)). In this configuration, each platinum (Pt) atom is covalently bonded to two chalcogen atoms (S or Se) and is octahedrally coordinated by six chalcogen atoms, as illustrated in Fig. S1(c) [47,53,73,74]. The unit cell of $PtX_2$ (X = S, Se) contains three atoms, resulting in nine phonon branches at the Γ-point of the Brillouin zone. The corresponding irreducible representation is given as: $\Gamma_{iredd.} = A_{1g} + E_g + 2A_{2u} + 2E_u$ [62,75,76]. Phonon modes with $A_{1}g$ symmetry correspond to non-degenerate out-of-plane vibrations of the chalcogen atoms, whereas modes with $Eg$ symmetry correspond to in-plane vibrations, as depicted in Fig. S1 of the supplementary material [53,75].

Figures 1(a) and 1(b) illustrate the temperature evolution of the Raman spectra for few-layer $PtS_2$ and $PtSe_2$. The spectral range spans ~ 5-380 cm$^{-1}$ (left panel) and 550-800 cm$^{-1}$ (right panel) for $PtS_2$, and ~ 5-260 cm$^{-1}$ for $PtSe_2$. A total of twenty-three phonon modes were identified in $PtS_2$, labelled as P1, P*, P2-P15, and M1-M7, while eight modes were observed in $PtSe_2$, designated as S1-S8, as shown in Fig. 1. To extract the phonon self-energy parameters i.e. peak position and full width at half maximum (FWHM), each Raman spectrum was fitted using a sum of Lorentzian functions. The fitting results are presented in Figs. S5 and S6 of the supplementary material for $PtS_2$ and $PtSe_2$, respectively. Both low-frequency (P1, P*, P2-P4; S1-S4) and high-frequency (P5-P11; S5-S8) phonon modes exhibit pronounced temperature dependence. As temperature increases, nearly all modes undergo a redshift and become more prominent compared to their low-temperature counterparts. In $PtS_2$, the weak features observed in the ranges ~ 70-240 cm$^{-1}$ (M1-M7) and ~ 550-800 cm$^{-1}$ (P12-P15) also evolve with temperature, as shown in Fig. 1(a). Notably, the P8 and P9 modes, having $A_{1g}$ symmetry, display Davydov splitting due to strong interlayer coupling [53], and their temperature evolution is shown in Fig. S2. Interestingly, the low-frequency mode P* begins to emerge around 150 K and intensifies with increasing temperature. Additionally, broad peaks appearing in the ~ 550-800 cm$^{-1}$ range may correspond to higher-order Raman processes.

### 3.1. Temperature dependance of the phonon modes

#### 3.1.1. Temperature dependance of the low frequency phonon modes

The low-frequency Raman modes are generally attributed to the interlayer interaction in two-dimensional materials. Raman spectroscopy is a useful technique to investigate low-frequency phonon modes to probe interlayer interactions in 2D materials as a function of the temperature and thickness [67,68,70,77]. These interlayer interactions are highly sensitive to external perturbations such as temperature, electric/magnetic fields, and mechanical strain, all of which can modulate the hybridization mechanism and thereby alter the interlayer coupling significantly [43,61,74,75,78-80]. Raman spectroscopy serves as a powerful tool for probing such interactions, particularly through the analysis of low-frequency phonon modes that provide insights into the nature and dynamics of interlayer coupling in these 2D materials [43,53,66-68,70].

In this study, temperature-dependent low-frequency Raman measurements were performed to probe interlayer interactions in the systems under investigation. Figures 2(a) and 2(b) present the temperature evolution of self-energy parameters i.e. peak position and FWHM over the temperature range of ~ 5-300 K for few-layered $PtS_2$ and $PtSe_2$, respectively. For the case of $PtS_2$, a new peak (P*) emerges above ~150 K, exhibiting temperature dependence that influences both the position and FWHM of the P1 mode. The frequency of mode P2 increases up to ~ 80 K, with a drop around ~ 80 K and then again increases till ~ 150K with a jump, followed by a decrease at higher temperatures. Its FWHM decreases with increasing temperature, marked by clear slope changes near ~80 K and ~150 K. Modes P3 and P4 show a decreasing frequency trend with temperature, also displaying transitions around ~80 K and ~150 K. FWHM of mode P3 decreases till ~ 80 K, then shows a sharp increase and decreases slowly with further increase in temperature up to ~150 K. After that it shows a sudden decrease and then becomes nearly constant till 300K. For P4, the FWHM remains nearly constant up to ~ 80 K and increases afterwards, with a distinct slope change near ~150 K.

For $PtSe_2$, frequency of mode S1 decreases with increasing temperature with change in slope around ~ 70 K and 240 K; while FWHM decreases sharply up to ~ 70 K and decreases with a further increase in temperature with a small change in slope around ~ 240 K. Frequency of mode S2 decreases with an increase in temperature up to ~ 70 K, with further increase in temperature it decreases slightly and starts to increase above ~ 240 K; while FWHM increases with an increase in temperature up to ~ 70 K and becomes nearly constant up to ~ 240 K and increases with further increase in temperature. Mode S3 frequency remains nearly constant up to ~ 240 K and decreases slightly afterwards. The FWHM reduces up to ~70 K, and becomes nearly constant with further increase in temperature up to ~ 240 K and decreases with further increase in temperature. For mode S4, frequency initially increases with increase in temperature up to ~ 70 K and decreases with further increase in temperature with a subtle change in slope around ~ 240 K. Its FWHM is nearly constant up to ~ 70 K and increases with increase in temperature with change in slope around ~ 240 K.

Figures 3(a) and 3(b) show the temperature dependence of the intensity of the low-frequency phonon modes P1, $P^*$, P2-P4, and S1-S4 for few layer $PtS_2$ and $PtSe_2$, respectively. For $PtS_2$, the intensity of mode P1 increases with temperature, and the emergence of P* above ~150 K notably influences P1 intensity. The intensity of modes P2-P4 increases with increase in temperature with a change in slope around ~ 80 K and 150 K. In $PtSe_2$, mode S1's intensity decreases up to ~70 K, rises until ~240 K, and then plateaus. S2 and S4 modes intensities increases with temperature, with distinct slope variations at ~70 K and ~240 K. For S2, intensity increases with increasing temperature with a change in slope at ~ 240 K.

The pronounced temperature dependence of the self-energy parameters for low-frequency phonon modes, along with distinct anomalies around ~80 K and ~150 K for $PtS_2$ and ~70 K and ~240 K for $PtSe_2$, points to the possibility of underlying phase transitions and associated exotic phenomena. The emergence of the low-frequency mode P* above ~150 K further highlights a temperature-induced modification in interlayer interactions, suggesting nontrivial coupling mechanisms at play. To fully elucidate these behaviours, we suggest more detailed measurements/characterization on both samples in the temperature regime.

**3.1.2. Temperature dependance of the high frequency phonon modes: Anharmonicity, Thermal expansion coefficient, and intensity**

Figures 4(a-b) and 4(c-d) display the temperature dependence of the self-energy parameters i.e. peak frequency ( $\omega$ ) and FWHM for high-frequency phonon modes in few-layer $PtS_2$ and $PtSe_2$ across the temperature range of ~ 5-300 K. In both materials, most phonon modes exhibit temperature-induced softening (redshift) and broadening, with notable anomalies observed near ~ 80 K and ~ 150 K for $PtS_2$, and ~ 70 K and ~ 240 K for $PtSe_2$. Raman spectroscopy serves not only as a tool for probing lattice dynamics but also for investigating electronic phases, including Fermi liquids and their associated transition temperatures [81,82]. We note that for $PtSe_2$, resistivity measurements reported by J.M. Salchegger [83] suggest the presence of a Fermi-liquid regime below ~ 80 K.

In $PtS_2$, the phonon mode frequencies for P6, P9, and P10 decreases with increasing temperature, with a change in slope near ~ 80 K and a subtle deviation around ~150 K. Mode P8 displays a nearly constant frequency up to ~80 K, followed by a linear decrease at higher temperatures, with a subtle deviation near ~ 150 K. The FWHM of mode P6 increases steadily with temperature, showing discernible slope changes around ~80 K. For modes P8 and P10, the FWHM increases slightly up to ~ 80 K, remains approximately constant between ~ 80-150 K, and rises again beyond ~ 150 K. In contrast, mode P9 shows an initial decrease in FWHM up to ~ 80 K, stabilizes between ~80-150 K, and then continues to decrease at higher temperatures. In $PtSe_2$, the phonon mode frequencies S5-S8 remain nearly constant up to ~70 K, followed by a linear decrease with rising temperature, showing a subtle deviation near ~ 240 K. The FWHM of modes S5 and S6 increase with increasing temperature, showing a change of slope around ~70 K. Mode S7 exhibits an initial increase in FWHM up to ~70 K, followed by a decrease at higher temperatures, with a change in slope near ~240 K. For mode S8, slope variation is not well-defined due to spectral fluctuations

In the context of 2D TMDCs, the temperature dependence of phonon mode frequencies can be primarily attributed to two mechanisms: (1) anharmonic effect ($\Delta\omega_{anh.}(T)$) and (2) thermal expansion of the lattice ($\Delta\omega_{latt.}(T)$). The combined influence of these effects on the phonon frequency as a function of temperature can be given as [63,84].

$$\Delta\omega(T) = \Delta\omega_{anh.}(T) + \Delta\omega_{latt.}(T) \tag{1}$$

The first term in Equation (1) accounts for frequency renormalization due to anharmonicity, typically understood via a three-phonon interaction process given as [85].

$$\Delta\omega_{anh.}(T) = \omega(T) - \omega_0 = A\left(1 + \frac{2}{(e^x - 1)}\right) \tag{2}$$

Similarly, the temperature-induced broadening of phonon modes (i.e., change in FWHM) is given as:

$$\Delta\Gamma(T) = \Gamma(T) - \Gamma_0 = B\left(1 + \frac{2}{(e^x - 1)}\right) \tag{3}$$

Here, $\omega_0$ and $\Gamma_0$ represent the phonon frequency and FWHM at 0 K, respectively. $x = \hbar\omega_0 / 2K_B T$ and the constants A and B correspond to the strengths of frequency shift and linewidth broadening. Figures 4(a)-4(d) show the experimental data fitted with equations (2) and (3) using solid red lines for $PtS_2$ in the range of ~80-300 K and $PtSe_2$ from ~70-300 K. The extracted constant parameters from these fits are summarized in Table I for both samples.

The second term in equation (1) corresponds to the thermal expansion of the lattice which is also known as the quasi-harmonic effect and the associated change in frequency is given as:

$$\Delta\omega_{latt.}(T) = \omega_0 \left\{ \exp\left[ -3\gamma \int_{T_0}^{T} \alpha(T) dT - 1 \right] \right\} \tag{4}$$

Where, $\gamma$ is the Gruneisen parameter and $\alpha(T)$ is linear thermal expansion coefficient (TEC). For simplicity, we have written the product of $\gamma$ and $\alpha(T)$ as:

$$\gamma\alpha(T) = a_0 + a_1 T + a_2 T^2 + .... \tag{5}$$

Here, $a_0$, $a_1$, and $a_2$ are constants (extracted values are listed in table S1, in supplementary information). For $PtS_2$ and $PtSe_2$, the Gruneisen parameter ( $\gamma$ ) for the phonon modes are not known. Hence, we have taken three different values for each mode i.e., $\gamma$ =1, 2, and 3.

Figures 5(a) and 5(b) illustrate the temperature dependence of the extracted linear thermal expansion coefficient, α(T), for the phonon modes P6 and P8-P10 for few-layer $PtS_2$ , in the temperature range of ~ 80-300 K; and modes S5, S6, and S8 for $PtSe_2$ , covering the temperature range of 70-300 K. For $PtS_2$, α(T) associated with mode P6 increases steadily with temperature. Mode P8 exhibits a decreasing trend up to ~150 K, followed by a linear increase beyond this point. In the case of P9, α(T) rises up to ~ 220 K before declining at higher temperatures. TEC for mode P10 shows a linear increase up to ~180 K, with an abrupt decrease thereafter. For $PtSe_2$, TEC for mode S5 increases linearly with temperature, while modes S6 and S8 display a non-monotonic behaviour i.e. initially decreasing up to ~130 K, followed by a sharp increase with further temperature rise. The sign and magnitude of α(T) reflect the material's thermomechanical response: positive values indicate lattice expansion, whereas negative values signify contraction. Quantifying thermal expansion coefficients in 2D materials is instrumental for strain engineering, heterostructure assembly, and nanoscale device fabrication [24,86]. The extraction of the TEC for few-layer of both $PtS_2$ and $PtSe_2$ may be helpful for the design and fabrication of the devices based on the Pt-based dichalcogenides.

In 2D materials, investigating the temperature dependence of the intensity of the phonon modes is very crucial. The intensity behaviour, influenced by parameters such as temperature, excitation energy, and sample thickness, can offer critical insight into interlayer coupling, phonon population dynamics, and emergent phase transitions [64,70,87-91]. Figures 6(a) and 6(b) display the temperature evolution of phonon mode intensities for the modes P5, P9-P11 for $PtS_2$ and S5-S8 for $PtSe_2$. In $PtS_2$, intensities of modes P5 and P10 increase with temperature, showing a slope change at ~ 80 K and ~ 150 K. Mode P9 exhibits an intensity increase up to ~ 150 K, with a change of slope at ~ 80 K, followed by a decrease at higher temperatures. Mode P11 displays a consistent decrease in intensity as temperature increases, with anomalies near ~ 80 K and ~ 150 K. In $PtSe_2$, mode S5 show a gradual intensity increase with increasing temperature with subtle slope variations near ~ 70 K and ~ 240 K. The intensity of the mode S6 increases sharply up to ~ 70 K, increases modestly until ~ 240 K, and then decreases with further increase in temperature. The intensity of the mode S7 increases with an increase in temperature up to ~ 70 K and with further increase with temperature start to decrease with a clear change in slope around ~ 240 K. The intensity of the mode S8 increases up to ~ 70 K, declines slightly until ~ 240 K, and then rises again at higher temperatures. The observed anomalies for both systems are observed for the temperature dependence of the intensity of the phonon modes and it's also showing the intensity of the phonon modes is also influenced by transition from fermi to non-fermi regime at temperature around ~ 70 K for the case of $PtSe_2$.

### 3.2. Polarization dependance of the phonon modes

Polarization-dependent Raman measurements provide information about the symmetry of the interlayer and intralayer modes in 2D and other materials [70,92,93]. The Raman tensors for the phonon modes with symmetries $E_g$ and $A_{1g}$ are listed in supplementary material [53]. The polarization dependence of phonon mode intensity can be understood using a semi-classical framework. Within this approach, the phonon scattering intensity is given as : $I \propto \left| e_s{}^T.R.e_i \right|^2$, where, $e_i$ and $e_s$ are the unit vectors representing the electric field directions of the incident and scattered light, respectively. R denotes the Raman tensor corresponding to the phonon mode under consideration, and T signifies matrix transposition. A matrix representation of the unit vectors pointing in the direction of the incident and scattered light is given as follows: $\hat{e}_i = [\cos(\theta+\theta_0) \quad \sin(\theta+\theta_0) \quad 0]$ and $\hat{e}_s = [\cos(\theta_0) \quad \sin(\theta_0) \quad 0]$. As illustrated in the schematic (Figure S1), $\theta$ represents the angle between the incident and scattered light polarization vectors, and $\theta_0$ is an arbitrary angle between the scattered light polarization and the x-axis. For phonons with $A_1g$ and Eg symmetries, the intensity variation with polarization angle $(\theta)$ is governed by their respective Raman tensors, which encode symmetry-dependent selection rules and is given as:

$$I_{A_{1g}} = a^2 cos^2(\theta) \quad , \quad \text{and} \tag{6}$$

$$I_{E_g} = \left| c\cos(\theta)(\cos^2(\theta_0) - \sin^2(\theta_0)) - 2c\sin(\theta)\sin(\theta_0)\cos(\theta_0) \right|^2 \text{, and} \tag{7(a)}$$

$$I_{E_g} = \left| c\sin(\theta)(\sin^2(\theta_0) - \cos^2(\theta_0)) - 2c\cos(\theta)\sin(\theta_0)\cos(\theta_0) \right|^2 \tag{7(b)}$$

Taking $\theta_0 = 0$, without loss of generality. Above equations 7(a) and 7(b) becomes:

$$I_{E_g} = c^2 \cos^2(\theta) \text{,} \quad \text{and} \tag{8(a)}$$

$$I_{E_g} = c^2 \sin^2(\theta) \tag{8(b)}$$

Adding the above two equations 8(a) and 8(b), it becomes,

$$I_{E_g} = c^2 \cos^2(\theta) + c^2 \sin^2(\theta) = c^2 \tag{9}$$

Figures 7 and 8 present the polarization-dependent intensity profiles of various phonon modes, with solid red lines indicating fits based on the corresponding analytical expressions. For $PtS_2$, Phonon mode P6 exhibits quasi-isotropic intensity behaviour, with intensity maxima occurring near ~ $30^0$ and ~ $210^0$, fitted with equation (9). Modes P8 and P9 display characteristic two-fold symmetry and are well-fitted using Equation (6). Notably, several low-frequency modes exhibit pronounced polarization dependence. Modes P1, P3, and P4 show clear two-fold symmetry, indicative of interlayer breathing vibrations. Mode P* and P2 demonstrates quasi-two-fold symmetry suggesting their nature as an interlayer shear mode. For $PtSe_2$, mode S5 reveals quasi-isotropic intensity behaviour with intensity maxima around ~ $30^0$ and $210^0$, again fitted via equation (9). Mode S6 shows two-fold symmetry and follows Equation (6). Mode S7 exhibits isotropic behaviour, whereas mode S8 displays quasi-isotropic characteristics. Among the low-frequency mode S1 shows two-fold symmetry suggesting it as an interlayer breathing mode. Mode S3 demonstrates a near two-fold symmetry profile. These polarization dependent signatures reflect the rich vibrational landscape and interlayer coupling dynamics inherent to layered transition-metal dichalcogenides.

A comprehensive polarization-dependent analysis of both interlayer and intralayer phonon modes enables the assignment of symmetry for low-frequency and high-frequency vibrational modes in few-layer $PtS_2$ and $PtSe_2$. The symmetry assignment based on the polarization-dependent study may be a valuable guide for heterostructure design and fabrication, where precise control over vibrational and electronic properties is essential.

**Conclusion**

In summary, our combined temperature and polarization dependent Raman investigations offer critical insights into the vibrational properties and symmetry assignments of interlayer and intralayer phonon modes in few-layer $PtS_2$ and $PtSe_2$. The observed phonon anomalies and the emergence of a potential Fermi to non-Fermi liquid transition signature near 70 K in $PtSe_2$ underscore the subtle interplay between thermal, electronic, and structural dynamics in these group-10 TMDs. Furthermore, the determination of thermal expansion coefficients and symmetry-resolved vibrational modes provides a robust foundation for engineering next-generation heterostructure devices, with implications for thermal management, strain modulation, and symmetry-driven electronic functionalities in layered 2D materials.

**Acknowledgements**

P.K. thanks SERB (Project no. CRG/2023/002069) for the financial support and IIT Mandi for the experimental facilities.

**Data availability statement**

All data that supports the findings of this study are included within the article and supplementary file.

**Figures:**

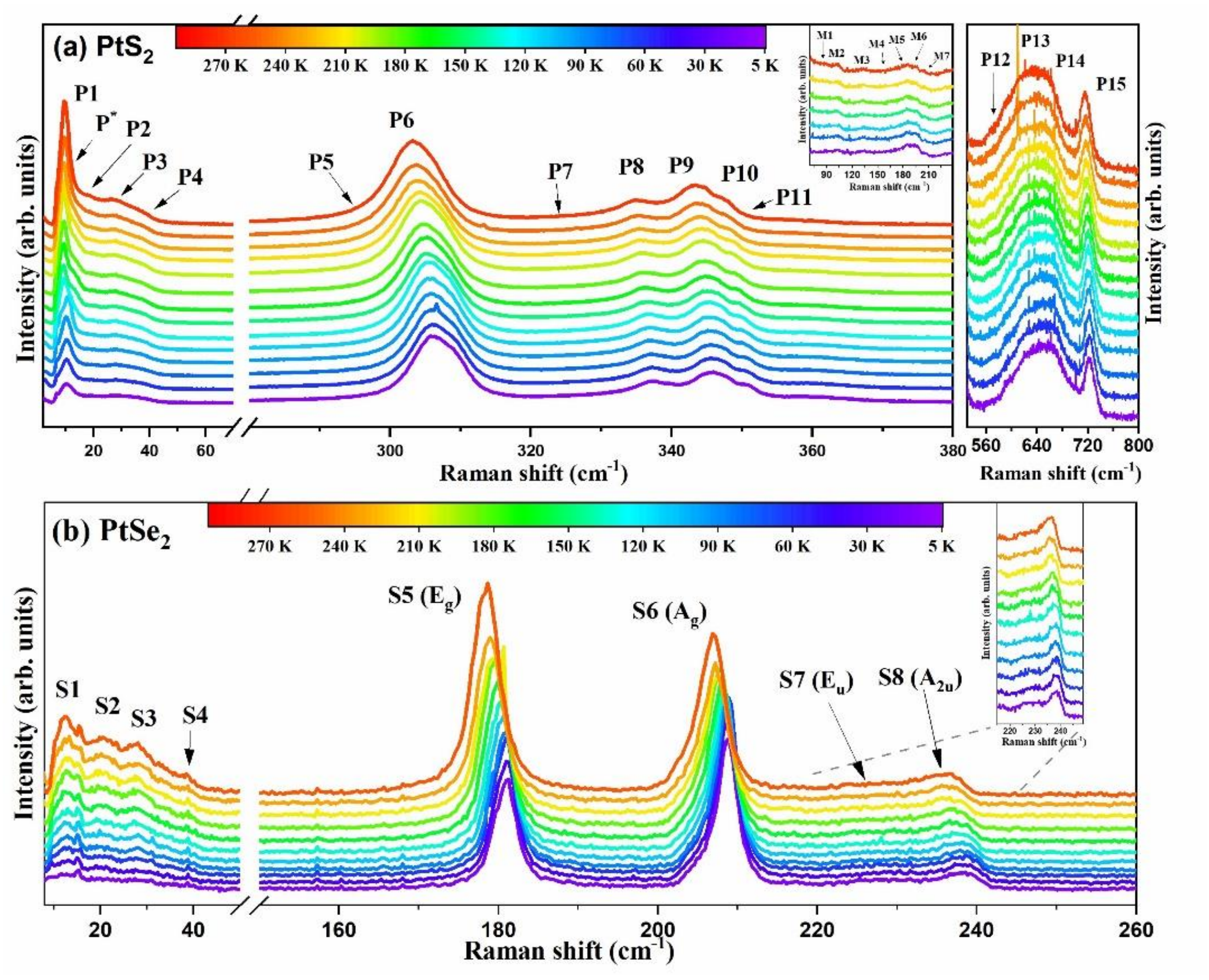

**Figure 1. (a)** Shows the evolution of the Raman spectrum for the few-layer PtS$_2$ in the frequency range of ~ 5-380 cm$^{-1}$ (left panel) and ~ 550-800 cm$^{-1}$ (right panel) in the temperature range of ~ 5-300 K. Inset (left panel) shows the temperature evolution of the weak spectrum in the frequency range of ~ 70-220 cm$^{-1}$. **(b)** Shows the temperature evolution of the Raman spectrum of few- layer PtSe$_2$ in the frequency range of ~ 5-260 cm$^{-1}$ in the temperature range of ~ 5-300 K. Inset shows the temperature evolution of the Raman spectrum in the frequency range of ~ 215-250 cm$^{-1}$.

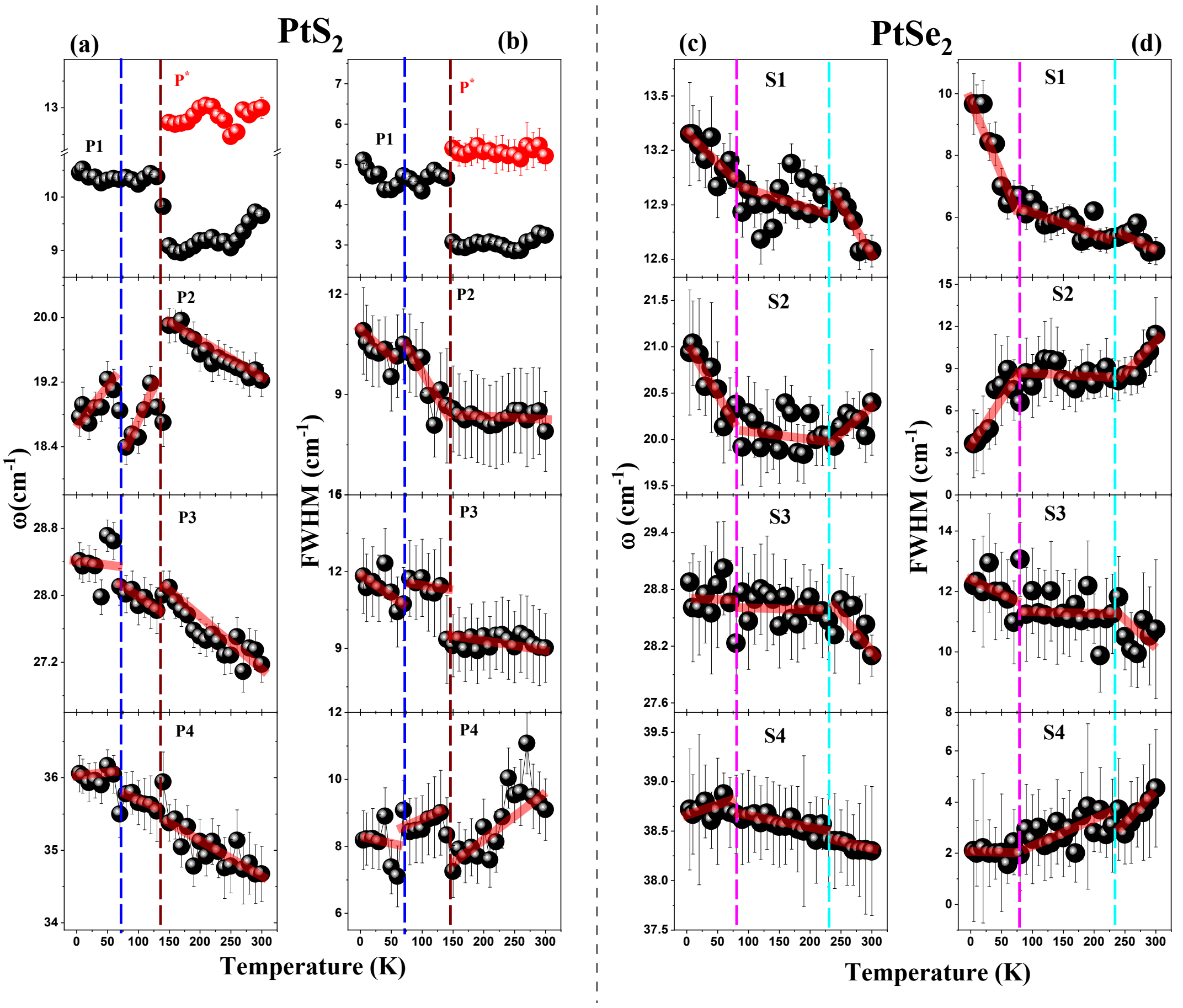

**Figure 2. (a)** and **(b)** Shows the temperature dependence of the peak position and FWHM of the low-frequency phonon modes in few-layer $PtS_2$. **(c)** and **(d)** Shows the temperature dependence of the peak position and FWHM of the low-frequency phonon modes in few-layer $PtSe_2$. The semi-transparent red lines are drawn as the guide to the eye.

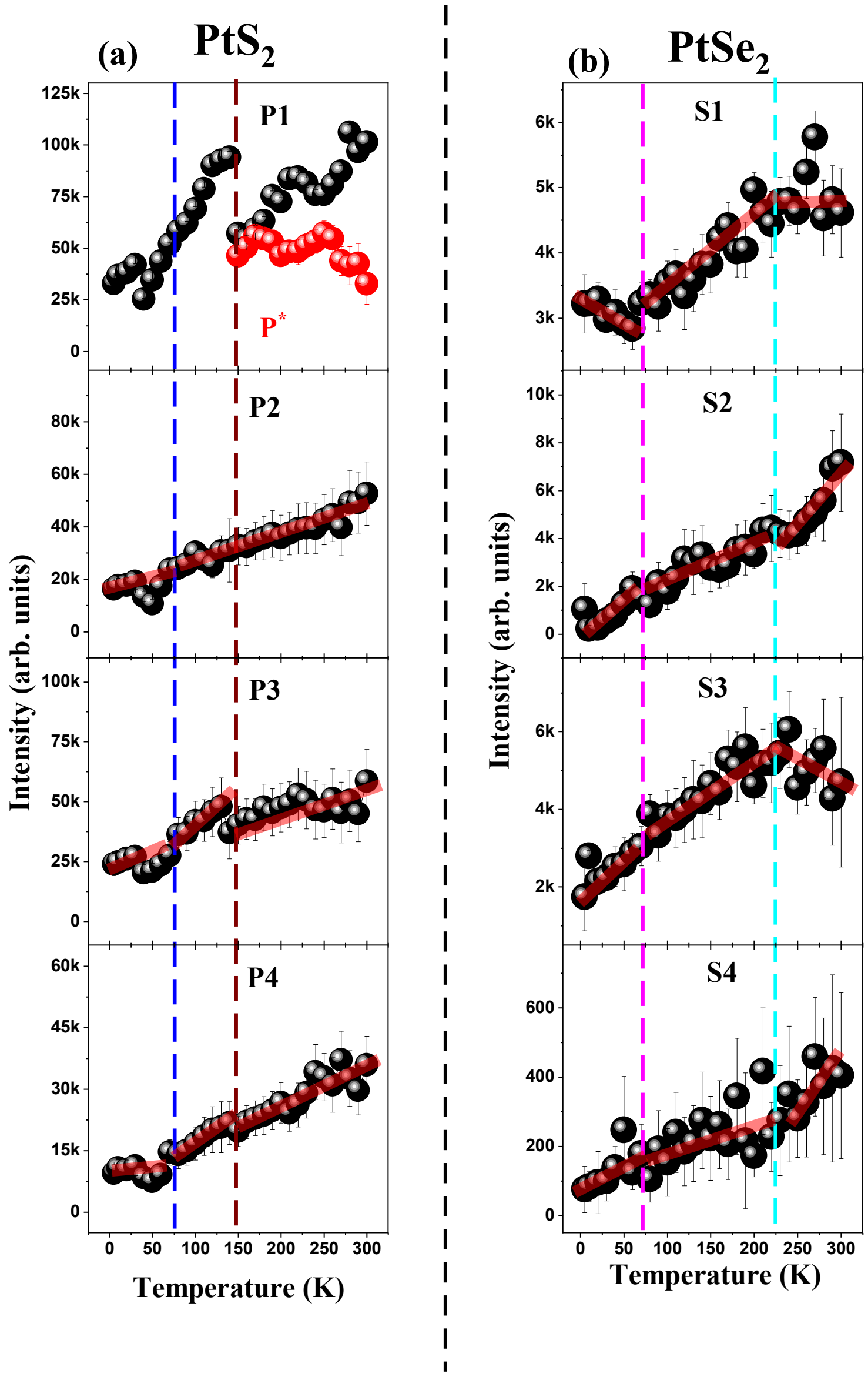

**Figure 3. (a)** and **(b)** Shows the temperature dependence of the intensity of the low-frequency phonon modes P1, P*, P2-P4 and S1-S4 for the few-layer $PtS_2$ and $PtSe_2$, respectively. Semi-transparent red lines are drawn as guide to the eye.

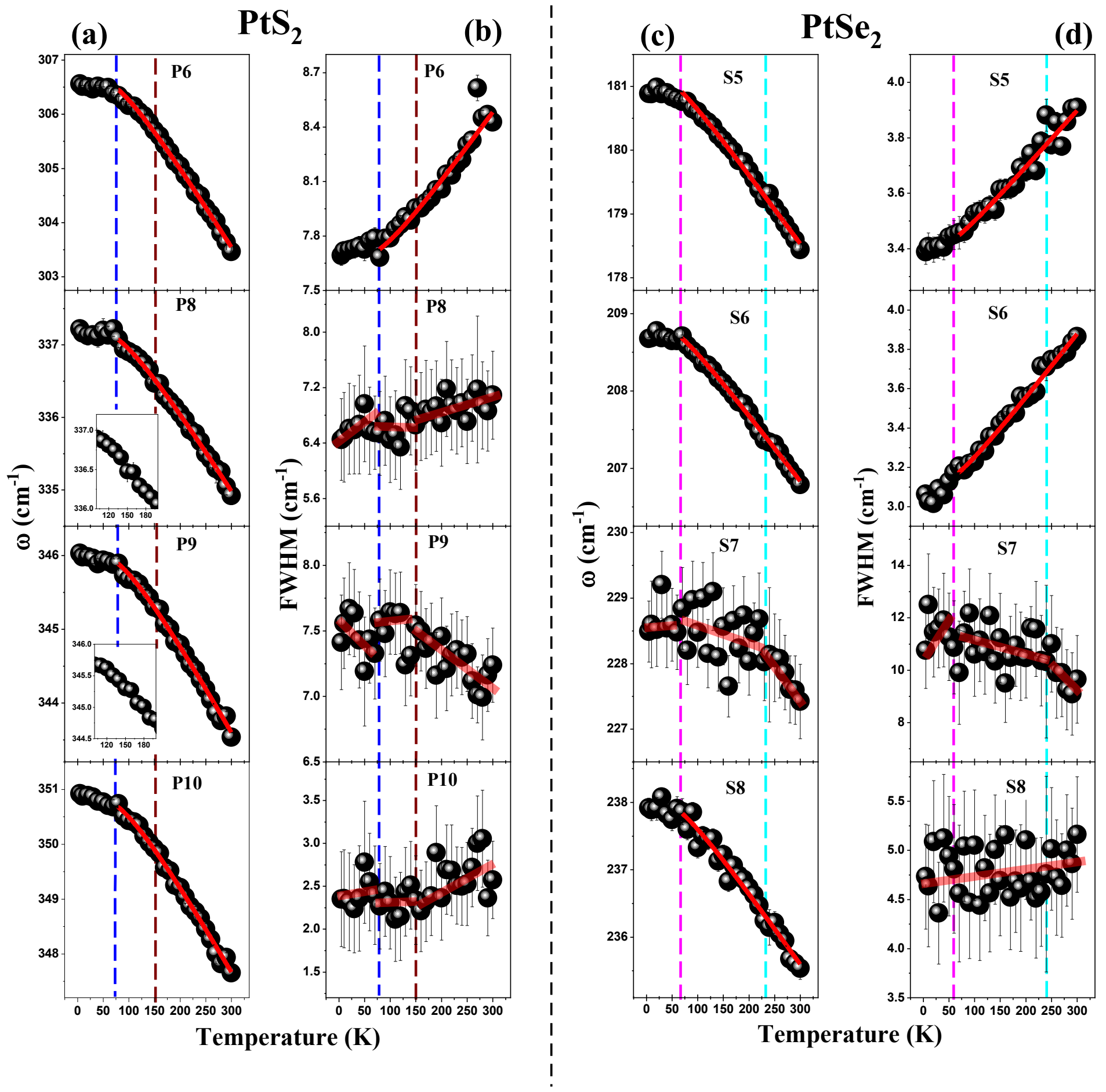

**Figure 4. (a)** and **(b)** Shows the temperature dependence of the peak position and FWHM of the phonon modes P6, P8-P10, respectively, for few-layer $PtS_2$. **(c)** and **(d)** Shows the temperature dependence of the peak position and FWHM of the phonon modes S5-S8, respectively, for few-layer $PtSe_2$. The solid red line shows a three-phonon fitting in the temperature range 80/70 to 300 K. The semi-transparent red lines are drawn as guide to the eye.

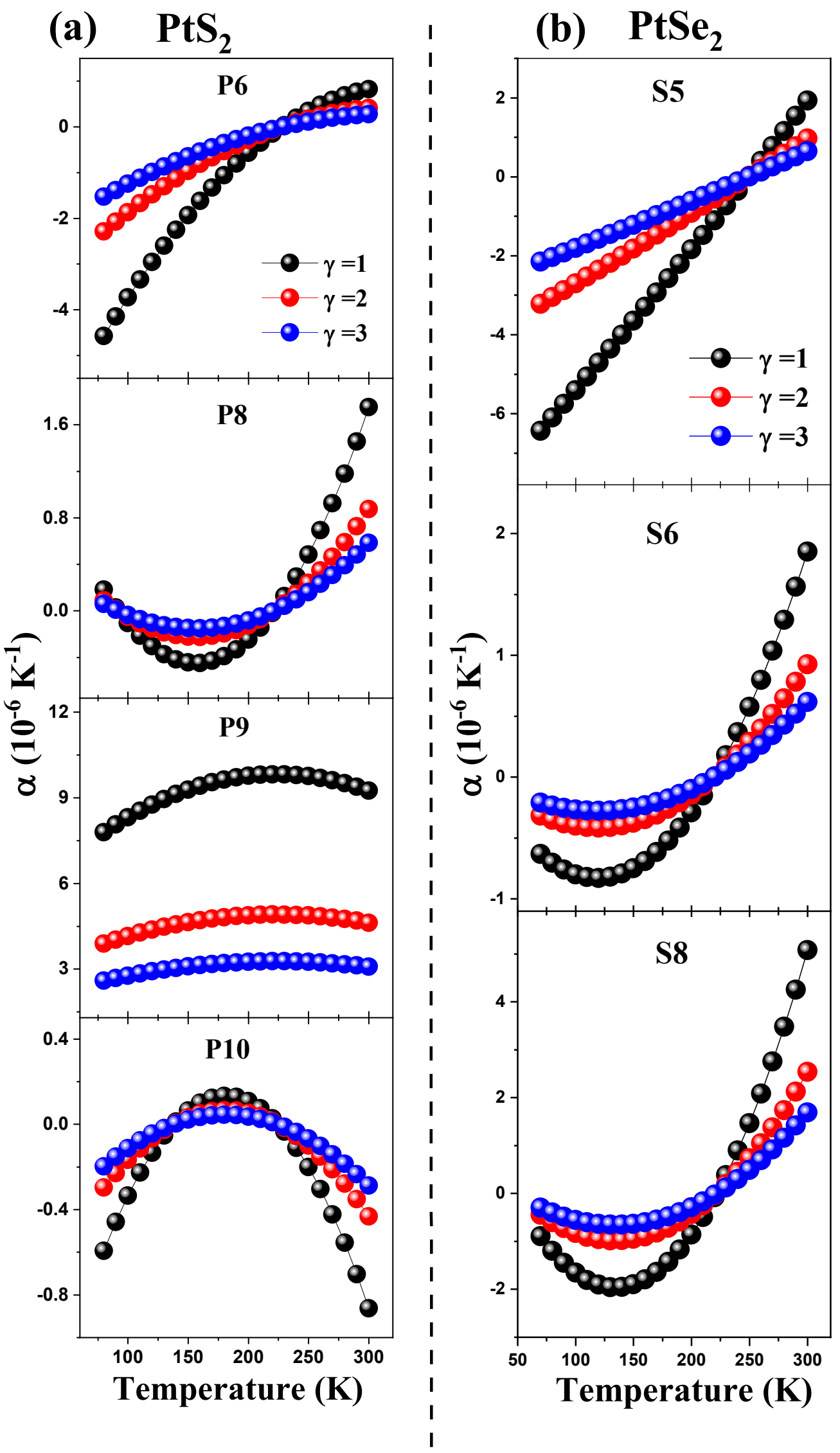

**Figure 5. (a)** and **(b)** Shows the temperature dependence of the thermal expansion coefficient (TEC) of the phonon modes P6, P8-P10 and S5, S6, S8 for the $PtS_2$ and $PtSe_2$ in the temperature range of ~ 80-300 K and 70-300 K, respectively.

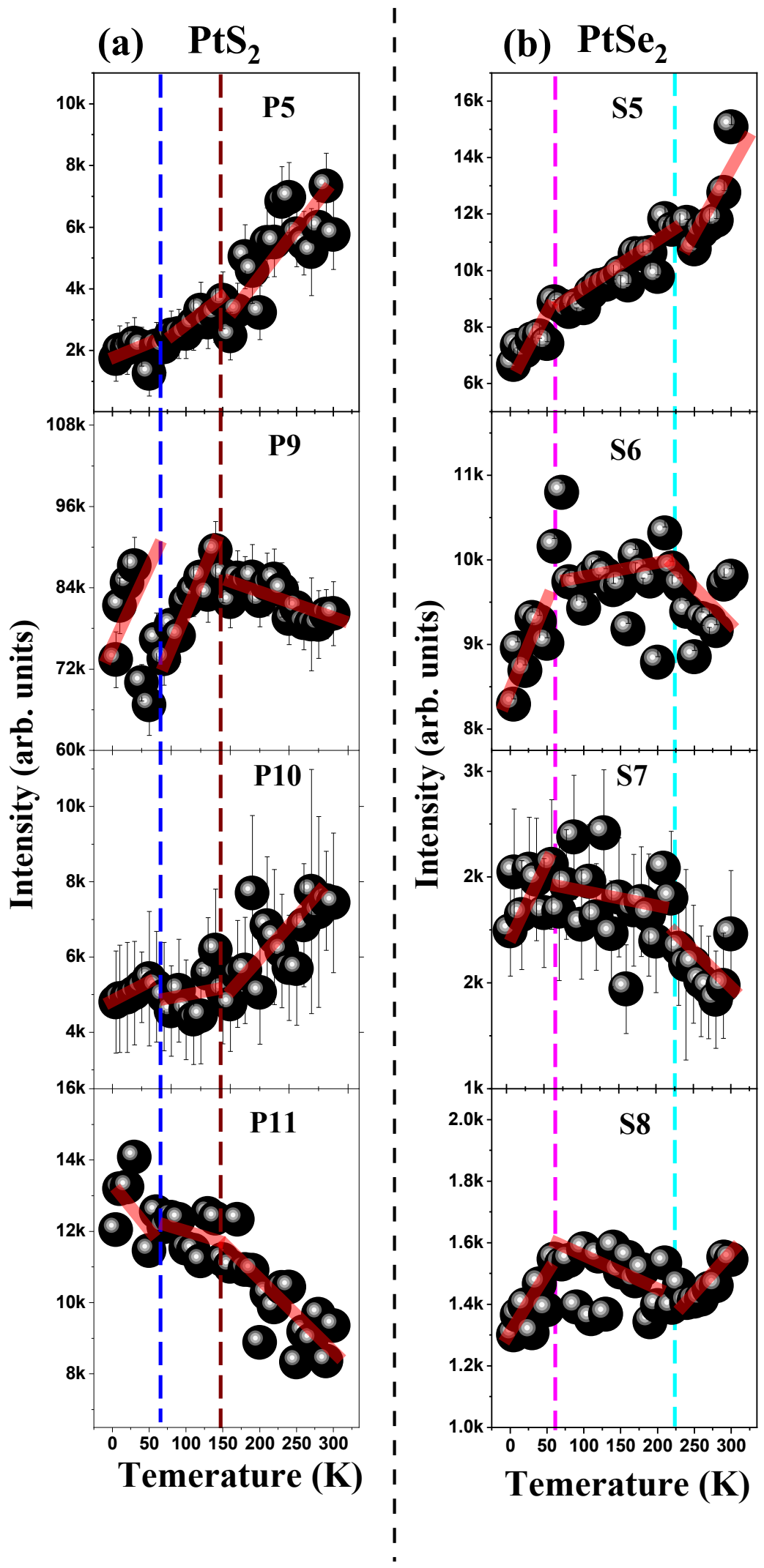

**Figure 6. (a)** and **(b)** Shows the temperature dependence of the intensity of the high-frequency phonon modes P5, P9-P11 and S5-S8 for the few-layer $PtS_2$ and $PtSe_2$, respectively. Semi-transparent red lines are drawn as guide to the eye.

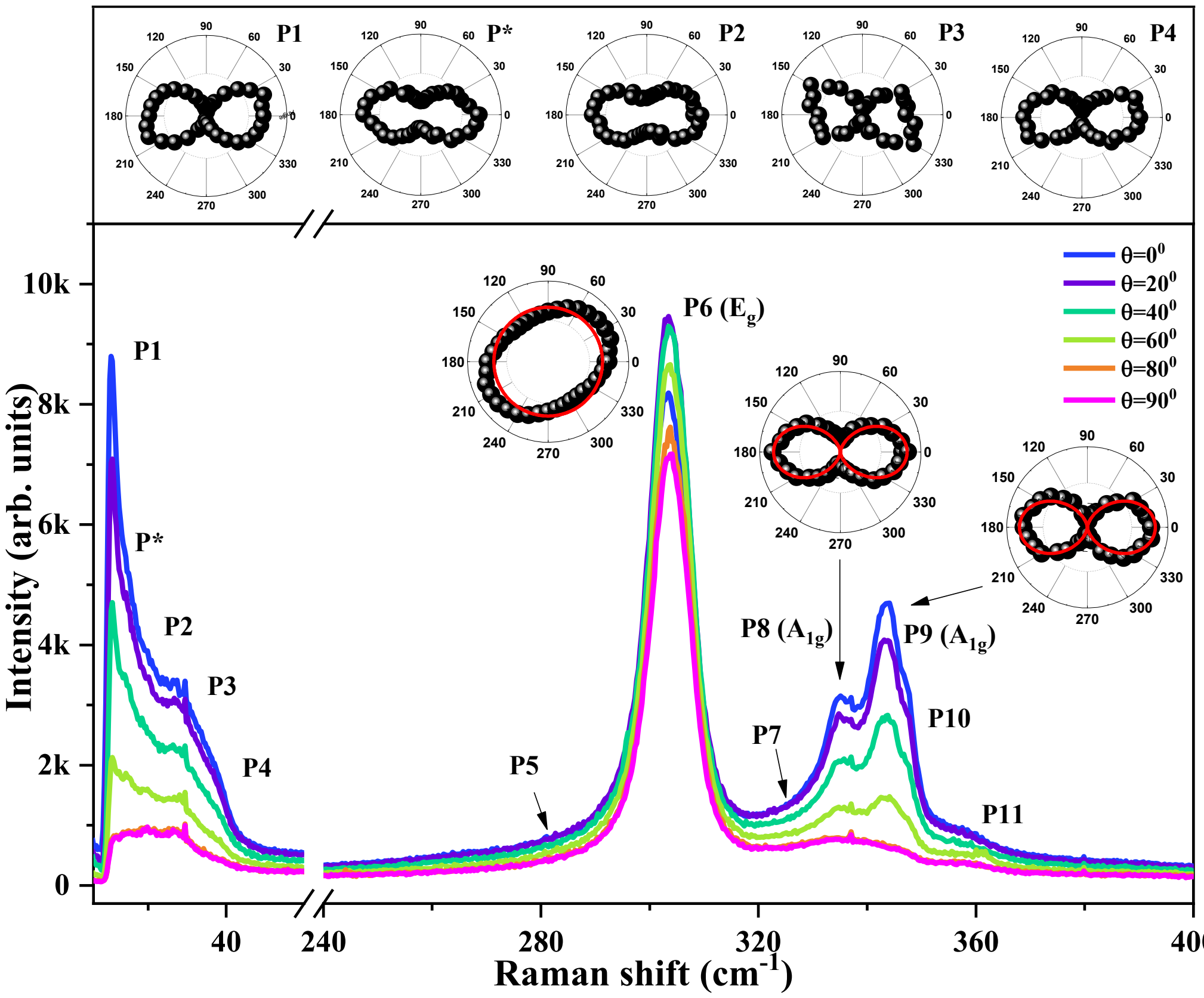

**Figure 7.** Shows the polarization dependence of the Raman spectrum in the frequency range of ~5- 400 $cm^{-1}$ for few layer $PtS_2$. Top panel shows the polar plots for the low frequency phonon modes. The solid red line shows the fitted curve with equations as described the text.

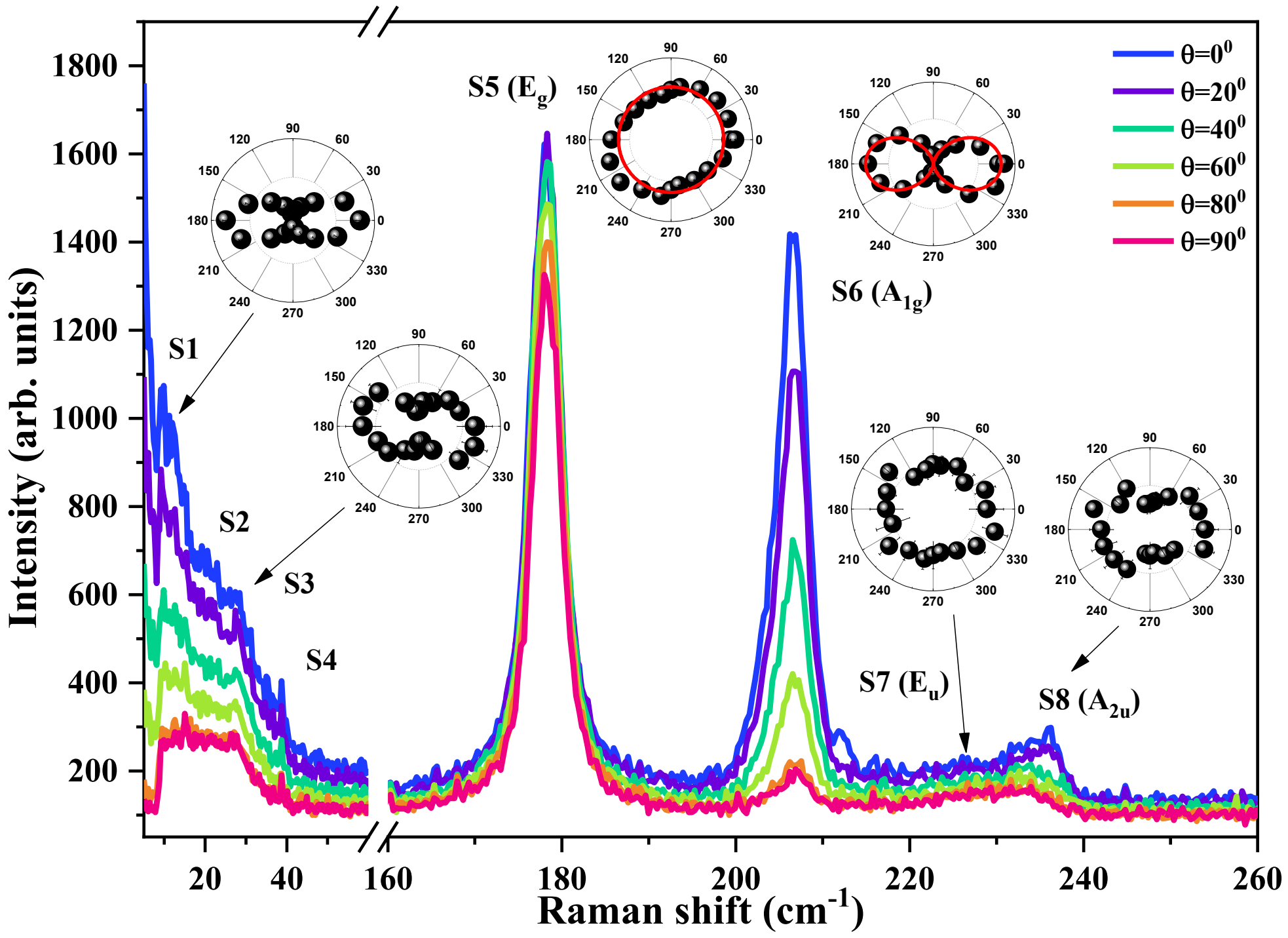

**Figure 8.** Shows the polarization dependence of the Raman spectrum in the frequency range of ~5- 260 $cm^{-1}$ for few layer $PtSe_2$. The solid red line shows the curve fitted with equations as described in the text.

**Table-I:** List of the fitting parameters corresponding to the observed phonon modes in $PtS_2$ and $PtSe_2$ , fitted using the three-phonon fitting model in the temperature range of ~80 to 300 K and 70 to 300 K, respectively. The units are in $cm^{-1}$.

| Sample | Peak | $\omega_0$ | A | $\Gamma_0$ | B |
|---|---|---|---|---|---|
| $PtS_2$ | P6 | 308.39 ± 0.05 | -1.71 ± 0.02 | 7.22 ± 0.04 | 0.45 ± 0.02 |
| | P8 | 338.60 ± 0.03 | -1.39 ± 0.02 | - | - |
| | P9 | 347.61 ± 0.04 | -1.58 ± 0.02 | - | - |
| | P10 | 352.98 ± 0.05 | -2.13 ± 0.03 | - | - |
| $PtSe_2$ | S5 | 181.89 ± 0.04 | -0.72 ± 0.01 | 3.26 ± 0.02 | 0.14 ± 0.01 |
| | S6 | 209.51 ± 0.02 | -0.66 ± 0.01 | 2.86 ± 0.02 | 0.25 ± 0.01 |
| | S8 | 238.94 ± 0.08 | -0.93 ± 0.03 | | |

Supplementary material:

# Phonon anomalies, Anharmonicity, and thermal expansion coefficient in few layered $PtX_2$ (X= S, Se): A temperature dependent Raman study

Atul G. Chakkar[1,*], Chaitanya B. Auti[1], Gaurav Bassi[2], Mukesh Kumar[2], and Pradeep Kumar[1,#]

[1]*School of Physical Sciences, Indian Institute of Technology Mandi, Mandi-175005, India*

[2]*Functional and Renewable Energy Materials (FREM) Laboratory, Department of Physics, Indian Institute of Technology Ropar, Rupnagar, Punjab, 140001, India*

**E-mail: atulchakkar16@gmail.com*

*#E-mail: pkumar@iitmandi.ac.in*

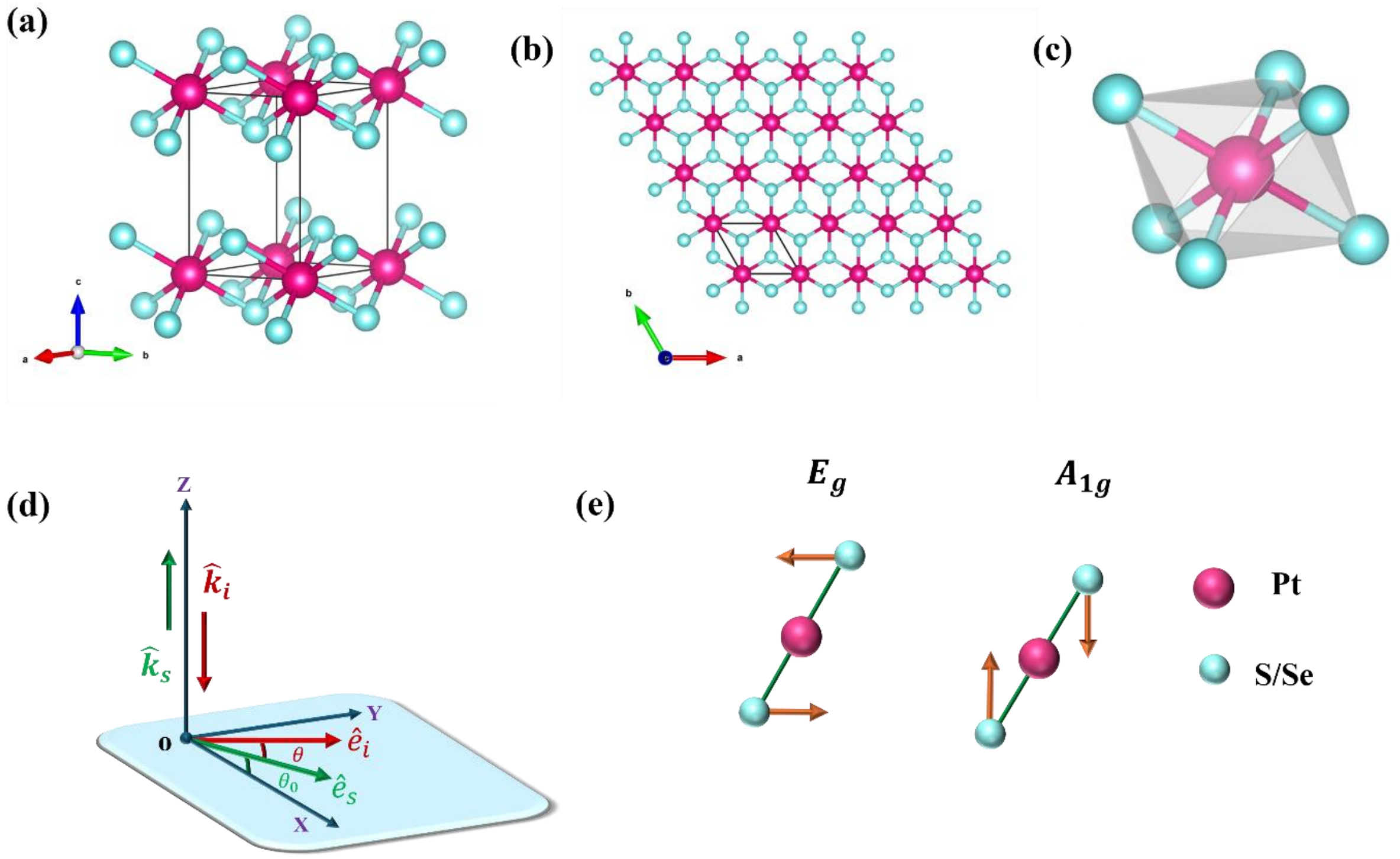

**Figure S1.** **(a)** Shows the primitive unit cell of 1T-$PtX_2$. **(b)** Shows the top view of the crystal structure. **(c)** Shows the Pt atom is octahedrally coordinated by chalcogen atoms (S, Se) in 1T-phase. **(d)** Shows the schematic for the polarization configuration. **(e)** Shows the ball-stick

diagram representing the lattice vibration corresponding to the in-plane ($E_g$) and out-of-plane ($A_{1g}$) Raman active modes.

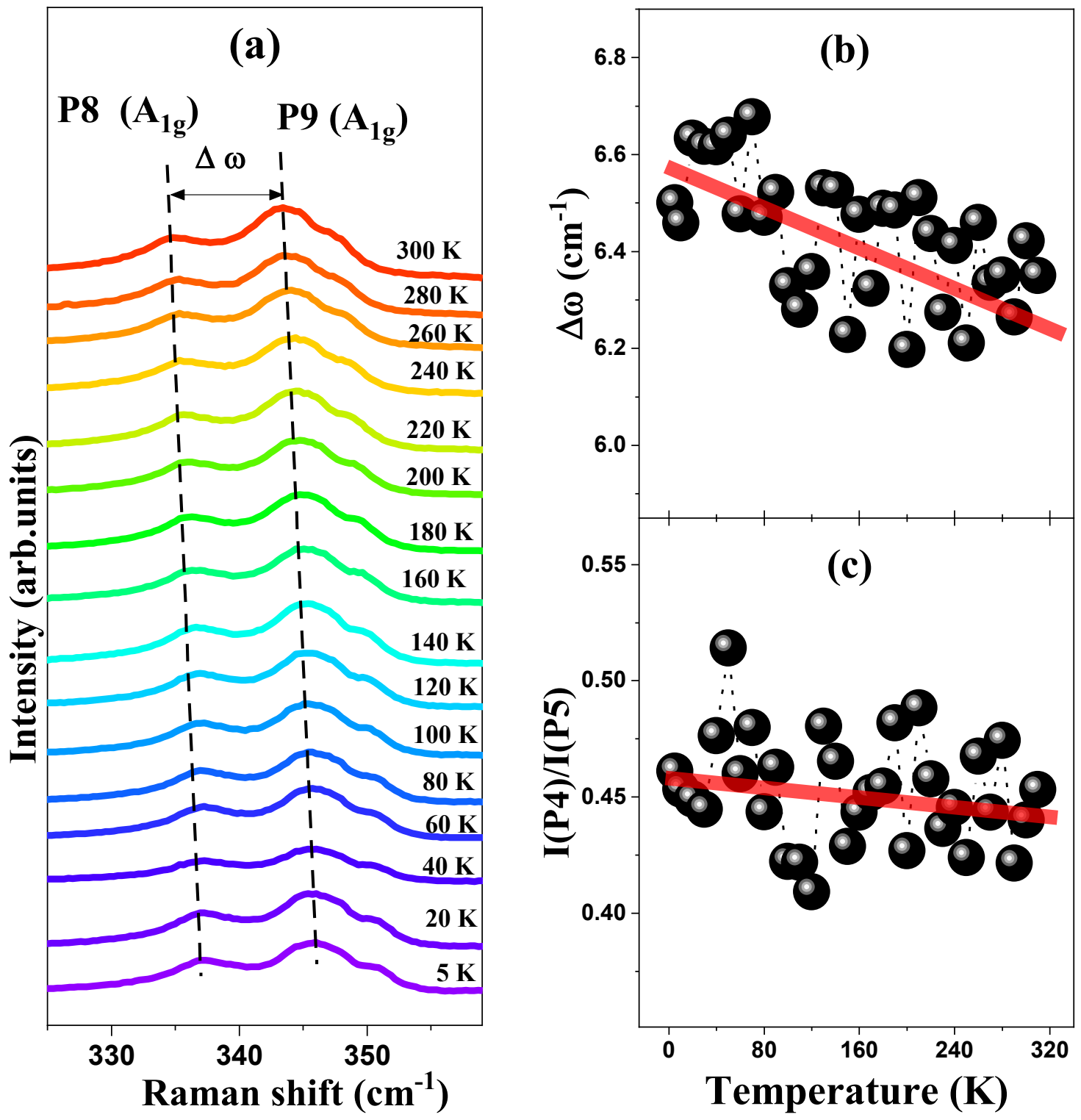

**Figure S2: (a)** Shows the temperature dependence of the Davydov splitting of intralayer $A_g$ mode in the temperature range of ~ 5-300 K, for few layer $PtS_2$. **(b)** Shows the temperature dependence of the separation between splitting (Δω). **(c)** Shows the ratio of the intensity of the modes P4 and P5 in the temperature range of ~ 5-300 K. The semi-transparent red lines are drawn for the guide to the eye.

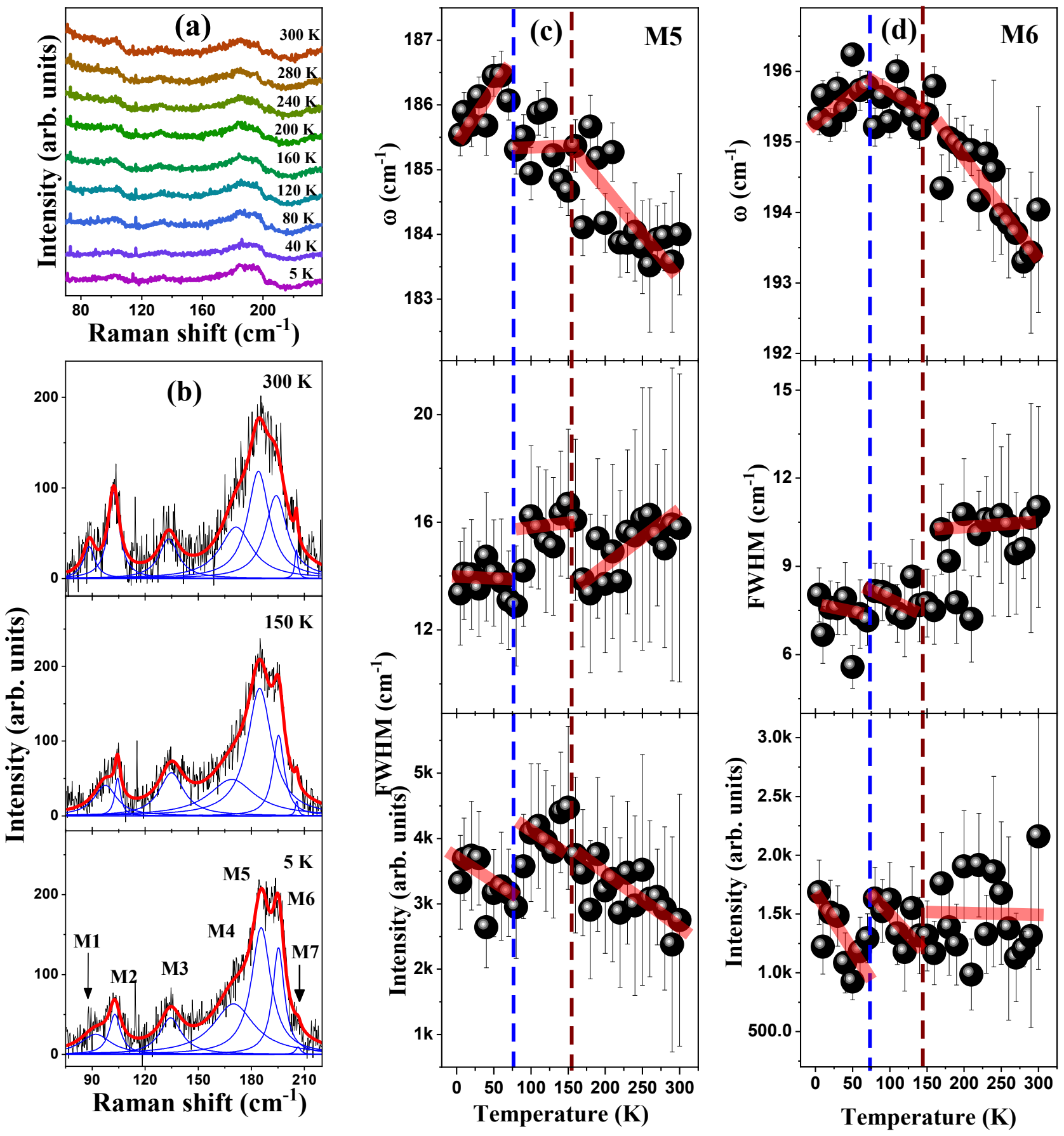

**Figure S3: (a)** Shows the temperature evolution of the Raman spectrum in the frequency range of ~ 70-240 $cm^{-1}$. **(b)** Shows the fitted Raman spectrum at three different temperatures ( i.e., 5,150, and 300 K). **(c)** and **(d)** shows the temperature dependence of the peak position, FWHM, and intensity of the peaks M5 and M6, respectively. The semi-transparent red lines are drawn for the guide to the eye.

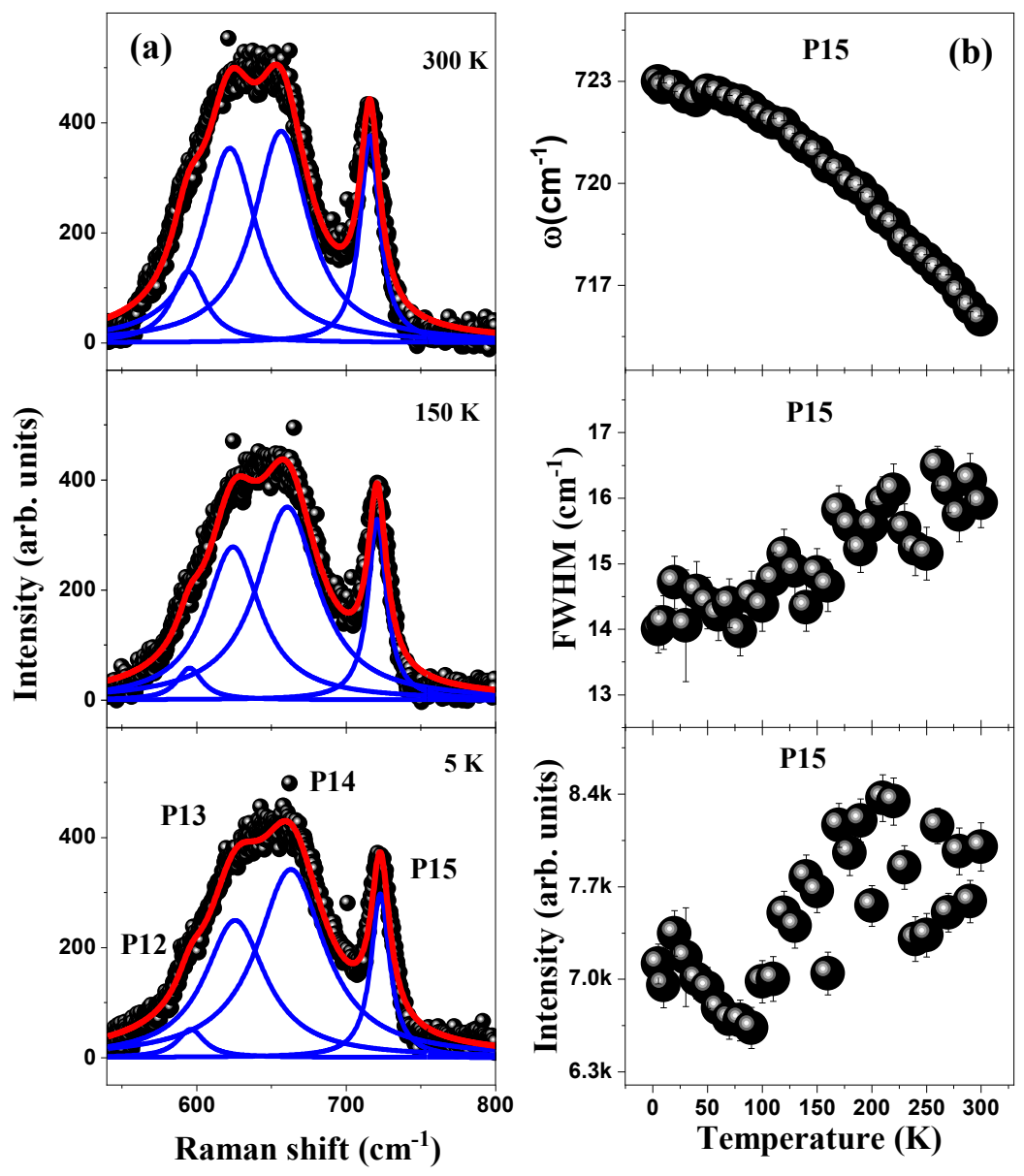

**Figure S4: (a)** Shows the fitted Raman spectrum at three different temperatures ( i.e., 5,150, and 300 K) in the frequency range of ~ 540-800 $cm^{-1}$ for $PtS_2$. **(b)** shows the temperature dependence of the peak position, FWHM, and intensity of the peak P16.

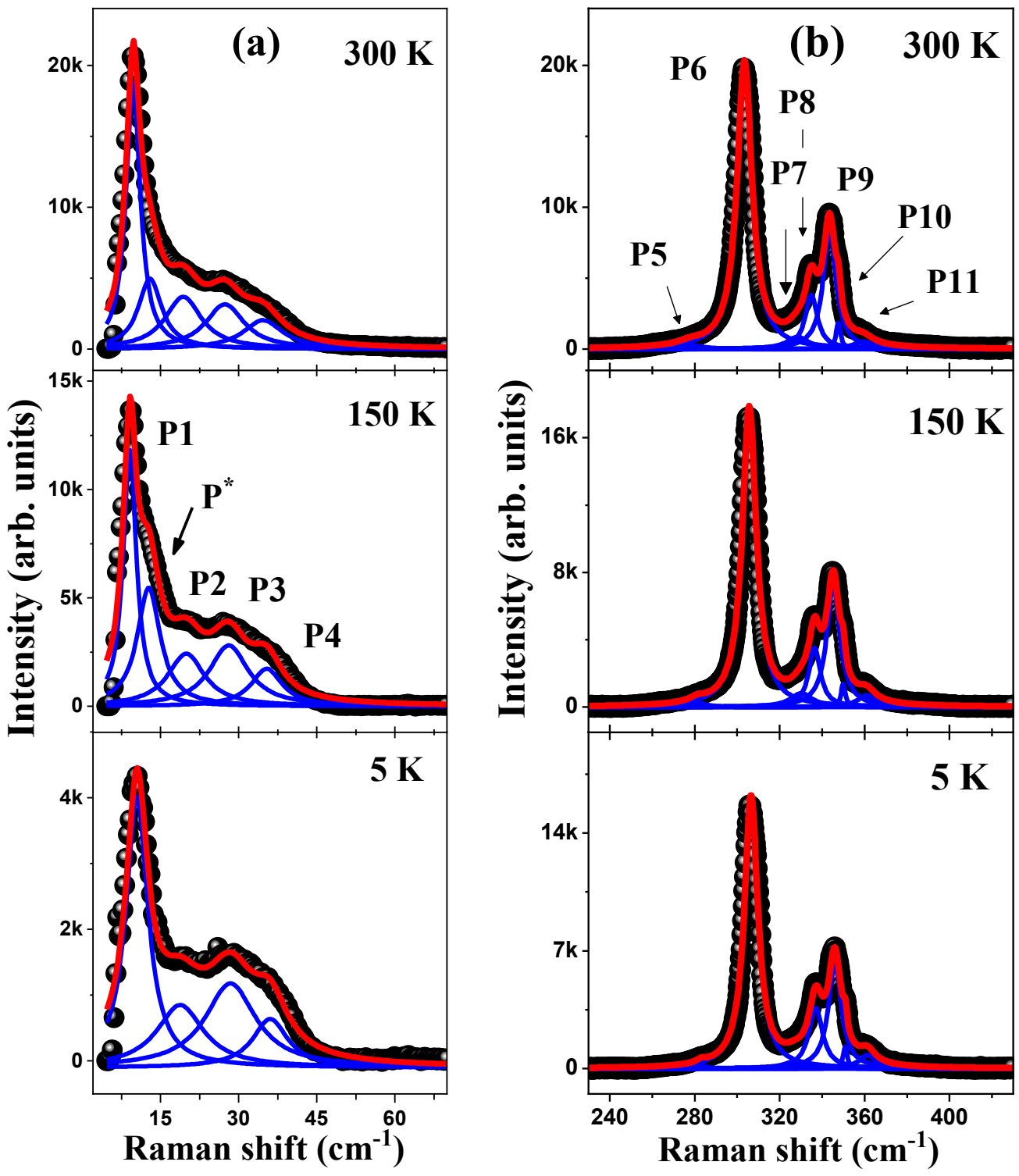

**Figure S5 :** Fitted Raman spectra for few-layer $PtS_2$ in the frequency range of **(a)** ~ 5-70 $cm^{-1}$ and **(b)** ~ 230-430 $cm^{-1}$ at three different temperatures 5, 150, and 300 K, respectively.

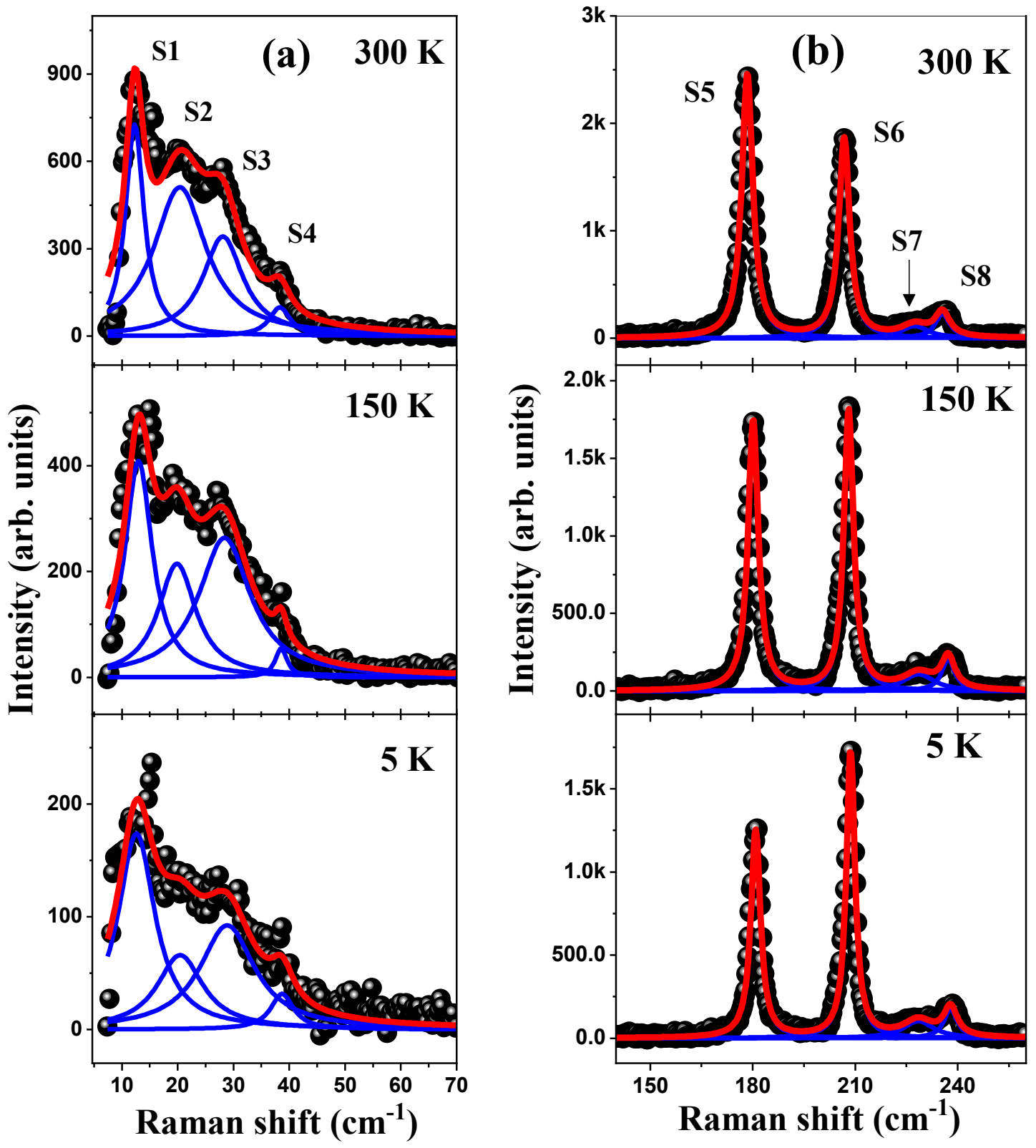

**Figure S6 :** Fitted Raman spectra for few-layer $PtSe_2$ in the frequency range of **(a)** ~ 5-70 $cm^{-1}$ and **(b)** ~ 140-260 $cm^{-1}$ at three different temperatures 5, 150, and 300 K, respectively.

❖ **Raman tensors for $PtX_2$ (X=S, Se):**

$$A_{1g} = \begin{pmatrix} a & 0 & 0 \\ 0 & a & 0 \\ 0 & 0 & b \end{pmatrix}, \text{ and}$$

$$E_g = \begin{pmatrix} c & 0 & 0 \\ 0 & -c & d \\ 0 & d & 0 \end{pmatrix}, \begin{pmatrix} 0 & -c & -d \\ -c & 0 & 0 \\ -d & 0 & 0 \end{pmatrix}$$

**Table-S1:** List of extracted constants with thermal expansion coefficient fitting along with three phonon anharmonicity.

| Sample | Peak | $a_0$ ($10^{-6}$) | $a_1$ ($10^{-8}$) | $a_2$ ($10^{-11}$) |
|---|---|---|---|---|
| $PtS_2$ | P6 | -8.67 | 5.82 | -8.86 |
| | P8 | 2.18 | -3.36 | 10.71 |
| | P9 | -4.89 | 4.41 | -9.86 |
| | P10 | -2.20 | 2.59 | -7.08 |
| $PtSe_2$ | S5 | -8.77 | 3.26 | 1.01 |
| | S6 | 0.33 | -1.94 | 8.17 |
| | S8 | 2.69 | -6.91 | 25.69 |